\documentclass[aps,pre,reprint,showpacs,floatfix,superscriptaddress]{revtex4-1}
\usepackage[colorlinks,linkcolor=red,anchorcolor=red,citecolor=blue,urlcolor=blue]{hyperref}
\usepackage{amsmath,txfonts,sidecap}
\usepackage{graphicx}
\usepackage{dcolumn,enumitem}
\usepackage{float}
\allowdisplaybreaks[3]

\begin{document}

\title{The nature of epidemic criticality in temporal networks}

\author{Chao-Ran Cai}
\address{School of Physics, Northwest University, Xi'an 710127, China}
\address{Shaanxi Key Laboratory for Theoretical Physics Frontiers, Xi'an 710127, China}
\author{Yuan-Yuan Nie}
\address{School of Physics, Northwest University, Xi'an 710127, China}
\author{Petter Holme}\email{petter.holme@aalto.fi}
\affiliation{Department of Computer Science, Aalto University, Espoo, Finland}
\affiliation{Center for Computational Social Science, Kobe University, Kobe, Japan}

\date{\today}
\begin{abstract}
Analytical studies of network epidemiology almost exclusively focus on the extreme situations where the time scales of network dynamics are well separated (longer or shorter) from that of epidemic propagation. In realistic scenarios, however, these time scales could be similar, which has profound implications for epidemic modeling (e.g., one can no longer reduce the dimensionality of epidemic models). We build a theory for the critical behavior of susceptible-infected-susceptible (SIS) epidemics in the vicinity of the critical threshold on the activity-driven model of temporal networks. We find that the persistence of links in the network leads to increasing recovery rates reducing the threshold. Dynamic correlations (coming from being close to infected nodes increases the likelihood of infection) drive the threshold in the opposite direction. These two counteracting effects make epidemic criticality in temporal networks a remarkably complex phenomenon.
\end{abstract}

\maketitle


The quest for epidemic models that are both realistic, analytically tractable, and realistic is a fundamental challenge to theoretical research. This is never as difficult as when one cannot rely on ignoring a structure, by assuming it constant. When building an analytical theory, how fast the network changes and how fast the epidemic spreads are two dynamics that one would like to reduce to one. From a medical point of view, that might not work---an incipient outbreak might sweep over the population in a matter of weeks, the same time scale of updates to the friendship network~\cite{fffng,bansal}. This is the motivation for temporal network epidemiology, where the networks do not necessarily change slower or faster than the disease propagation~\cite{masuda_holme,masudaholme_book}.

There are three main philosophies of how to go beyond static network epidemiology to include the temporality of edges. The first approach temporal contacts as possible contagion events~\cite{PhysRevE.94.022305}. This approach often relies on empirical data of the contacts and analyzes them by computer simulations. The second approach, \textit{adaptive networks}, models the contacts without real-world data but assumptions about how they change given the state of the disease~\cite{PhysRevLett.96.208701}. In this paper, we will take a third approach which is also purely model-based, trying to emulate dynamic contact structures that real epidemics spread on, while being tractable for analytical calculations. Typically these models---such as blinking networks~\cite{doi:10.1137/120893409} and activity-driven networks~\cite{Perra2012,PhysRevLett.112.118702}---assume an underlying static network over which they generate active contacts.

The original activity-driven network model was proposed by Perra et al.~\cite{Perra2012}, as follows: (i) The network is initialized to be completely disconnected and an activity $a_i=\eta x_i$ is assigned to each node $i$ based on the given activity distribution $F(x)$; (ii) At each discrete time $t$, each node $i$ becomes active with probability $a_i\Delta t$ and randomly creates $m$ undirected links with other nodes. (iii) The epidemic dynamics take place over the instantaneous network; (iv) At the next time step $t+\Delta t$, all the links are removed, and the process resumes from step (ii).
The activity-driven network can be considered as the simplest yet nontrivial framework to study dynamical processes unfolding on temporal networks~\cite{Ribeiro2013}. Only one variable, $F(x)$, is time-invariant and represents the degree distribution of an aggregated network~\cite{Perra2012,PhysRevLett.112.118702,PhysRevE.87.062807,Karsai2014,Ribeiro2013}.
Over the past decade, the epidemic dynamics on activity driven networks have extended in different directions~\cite{PhysRevE.90.042801,PhysRevX.5.021005,Sun2015,PhysRevLett.117.228302,RIZZO2016212,Ubaldi2016,PhysRevE.96.042310,PhysRevE.98.062315,PhysRevE.98.062322,PhysRevE.97.012313,8566006,doi:10.1098/rsif.2020.0875,PhysRevE.104.014307,PhysRevE.104.044307}, including tackling real epidemiological models~\cite{RIZZO2016212,doi:10.1098/rsif.2020.0875}, the effects of heterogeneity~\cite{PhysRevE.90.042801,PhysRevE.104.014307,Ubaldi2016}, and the introduction of network features such as memory effects~\cite{PhysRevE.98.062315,Sun2015} and individual attractiveness distribution~\cite{PhysRevE.96.042310,PhysRevE.104.044307}.

In static network epidemiology, the canonical models (SIR and SIS) are effectively governed by one parameter $\beta/\mu$---the ratio of infection to recovery rate. Therefore, most studies of activity-driven networks also focused on this ratio~\cite{Perra2012,PhysRevE.90.042801,Sun2015,PhysRevE.96.042310,PhysRevE.98.062315,8566006,doi:10.1098/rsif.2020.0875,PhysRevE.104.014307,PhysRevE.104.044307}. This approach, effectively studying the model in the limit of fast network dynamics, goes against the idea of temporal networks as the modeling framework for intermediate time scales. In this limit, the $\beta/\mu$-based mean-field approach accurately describes the criticality of compartmental models on activity-driven networks. However, reality can have slower network dynamics, and then the mean-field analysis will fail.

More precisely: The thresholds of both annealed (the limit of rapidly changing) and static (the limit of slowly changing) networks depend only on $\beta/\mu$, but their actual critical values $(\beta/\mu)_c$ are different because of dynamical correlations in the static networks~\cite{PhysRevLett.116.258301}. (I.e., infected nodes tend to group together.) Since the activity-driven network can interpolate between these extremes, by keeping one of $\beta$ or $\mu$ fixed, we can understand that tuning only the other must change the threshold (contradicting the assumption that $\beta/\mu$ fully describes the criticality of epidemics on the activity-driven model).

In this paper, we employ a continuous time description~\cite{Leung2017,PhysRevLett.117.228302} of the coevolution of the network and SIS process. We use simulations to see the effects of both network correlations (reflecting the persistence of social contacts) and dynamic correlations caused by the inertia of the disease propagation. We complement these simulations with theoretical calculations that, for the first time in the literature, include the above-mentioned network correlations but ignore dynamic correlations. The comparison of these two types of results allows us to see the effects of the two types of correlations.


Next, we turn to a technical definition of the model---a continuous time description of the coevolution of the network and SIS dynamics.
For the evolution of the network, each inactive (U) individual $i$ is activated with a rate $a_i=\eta x_i$, where $x_i$ is drawn from a probability distribution $F(x)$. 
When an individual gets activated, it randomly selects $m$ individuals to generate links to.
Active (A) individuals become unactivated with a rate $b$, and then remove all links to themselves.
Here, the use of U instead of I for inactive individuals is to avoid confusion with the concept of the epidemic model.
For the evolution of disease, we employ the classical SIS model. 
Susceptible (S) individuals become infected via contacts with infected (I) individuals at rate $\beta$ times the number of S-I links.
Infected individuals recover to susceptible with rate $\mu$.
The detailed simulation procedure see Appendix A.

 
We start by analyzing the network evolution, which is independent of the SIS model. 
Define $\rho^{A_a}$ as the fraction of active individuals with activity rate $a$, and $\rho^{A}=\int\rho^{A_a}da$ as the total fraction of active individuals. The dynamic equation of the fraction of active individuals of class $a$ in the network can be written as
\begin{equation}\label{eq1}
\frac{d\rho^{A_a}}{dt}=a(1-\rho^{A_a})-b\rho^{A_a}.
\end{equation}
Here, the first term of Eq.~\eqref{eq1} represents the spontaneous creation of active individuals, while the second term represents the spontaneous annihilation.
When the evolution of the network reaches its steady state, i.e. $\frac{d\rho^{A_a}}{dt}=0$, we have
\begin{equation}\label{eq2}
\rho^{A_a}=\frac{a}{a+b},\ \ \rho^{A}=\left\langle \frac{a}{a+b}\right\rangle,
\end{equation}
where $\left\langle \frac{a}{a+b}\right\rangle=\int\frac{\eta xF(x)}{\eta x+b}dx$.
We define $\langle k_{A,a}\rangle$ as the average degrees of active individuals in class $a$.
Considering that the links created per unit time in the network system should be equal to the disconnected ones, we have
\begin{equation}\label{eq3}
\int maN(1-\rho^{A_a})da=\int bN\rho^{A_a}\langle k_{A,a}\rangle da.
\end{equation}
Combining Eqs.~\eqref{eq2} and \eqref{eq3}, we obtain
\begin{equation}\label{eq4}
\langle k_A\rangle=\langle k_{A,a}\rangle=m,
\end{equation}
where $\langle k_{A}\rangle$ is the average degrees of active individuals.
Equation \eqref{eq4} also implies that the average degrees of inactive individuals in different classes are the same, and the average degree of inactive individuals is represented by $\langle k_U\rangle$.

Similarly, the number of links between active and inactive individuals is stable when the system reaches its steady state. 
Since $\int aN(1-\rho^{A_a})da$ individuals are activated per unit time, the number of new links to the inactive individuals is $m(1-\rho^{A})\int aN(1-\rho^{A_a})da$.
On the one hand, links are disconnected when the active node of the link deactivates.
On the other hand, when an inactive node in the link is activated, the class of the link changes.
Therefore, the amount of link reduction between active and inactive individuals per unit time is $\langle k_U\rangle \int N(1-\rho^{A_a})(a+b)da$.
Note that the number of links between active and inactive individuals is equal to $\langle k_U\rangle\int N(1-\rho^{A_a})da$, because there are no links between inactive and inactive individuals.
According to the relation $m(1-\rho^{A})\int aN(1-\rho^{A_a})da=\langle k_U\rangle \int N(1-\rho^{A_a})(a+b)da$, we have $\langle k_U\rangle=m\langle \frac{a}{a+b}\rangle(1-\langle \frac{a}{a+b}\rangle)$. By the relation $\langle k\rangle=\langle k_A\rangle \rho^{A}+\langle k_U\rangle(1-\rho^{A})$, the average degree of the network is
\begin{equation}\label{eq5}
\langle k\rangle=m\left\langle \frac{a}{a+b}\right\rangle\left(1+\left\langle \frac{b}{a+b}\right\rangle^2\right).
\end{equation}

Next, we consider the coevolution of networks and epidemics. Since individuals are infected through links, the evolution of the disease cannot be separated from the evolution of the network.
The individuals are classified as $XY$, where $X\in\{U,A\}$ and $Y\in\{S,I\}$. 
We introduce the notation $\rho^{XY_a}$ for the fraction of individuals with state $XY$ and class $a$; $\phi^{XY_a,X'Y'_{a'}}$ is the probability that an individual with state $XY$ and class $a$ is connected to an individual with state $X'Y'$ and class $a'$.
Note that $\phi^{US_a,US_{a'}}$, $\phi^{UI_a,UI_{a'}}$, and $\phi^{US_a,UI_{a'}}$ would be impossible.

From the above, we can state the master equations as follows, 
\begin{subequations}
\begin{eqnarray}\label{eq6}
\frac{d\rho^{UI_a}}{dt}&=&\beta\int\phi^{US_a,AI_{a'}}da'-(a+\mu)\rho^{UI_a}+b\rho^{AI_a},\\
\frac{d\rho^{AI_a}}{dt}&=&\beta\int\left(\phi^{AS_a,AI_{a'}}+\phi^{AS_a,UI_{a'}}\right)da'+a\rho^{UI_a}\nonumber\\
&&-(b+\mu)\rho^{AI_a}.
\end{eqnarray}
\end{subequations}
Since only active individuals can make connections and inactive individuals cannot, we can  approximate $\phi^{XY_a,X'Y'_{a'}}$ in Eq.~\eqref{eq6} in the form of $\rho^{XY_a}$ using the mean field anzats, i.e.
\begin{subequations}
\begin{align}\label{eq7}
\phi^{AS_a,AI_{a'}}= & 2\langle k_A\rangle\rho^{AS_a}\rho^{AI_{a'}},\\
\phi^{AS_a,UI_{a'}}= & \langle k_A\rangle\rho^{AS_a}\rho^{UI_{a'}},\\
\phi^{US_a,AI_{a'}}= & \langle k_A\rangle\rho^{US_a}\rho^{AI_{a'}}.
\end{align}
\end{subequations}
Note that the approximation of Eq.~\eqref{eq7} means that we ignore the correlation of dynamic.
By inserting Eq.~\eqref{eq7} into Eq.~\eqref{eq6}, and then integrating both sides of Eq.~\eqref{eq6} with respect to $a$, we get
\begin{subequations}
\label{eq8}
\begin{eqnarray}
\frac{d\rho^{AI}}{dt} = & \left(2\beta m\left\langle\frac{a}{a+b}\right\rangle-b-\mu\right)\rho^{AI}-2\beta m\left(\rho^{AI}\right)^2\nonumber\\
&+\left(\beta m\left\langle\frac{a}{a+b}\right\rangle+\langle a\rangle\right)\rho^{UI}-\beta m\rho^{UI}\rho^{AI},\\
\frac{d\rho^{UI}}{dt} = &\left(\beta m\left\langle\frac{b}{a+b}\right\rangle+b\right)\rho^{AI}-\left(\langle a\rangle+\mu\right)\rho^{UI}\nonumber\\
&-\beta m\rho^{UI}\rho^{AI},
\end{eqnarray}
\end{subequations}
where $\rho^{UI}=\int\rho^{UI_a}da$ and $\rho^{AI}=\int\rho^{AI_a}da$. See Appendix C for detailed derivations.

Now we perform a linear stability analysis for Eq.~\eqref{eq8} around the fixed point $\rho^{AI}=\rho^{UI}=0$, and after ignoring the higher order terms, the Jacobian matrix of Eq.~\eqref{eq8} can be written as
\begin{eqnarray}\label{eq9}
J=\left(\begin{array}{cccc}
2\beta m\left\langle\frac{a}{a+b}\right\rangle-b-\mu  &  \beta m\left\langle\frac{a}{a+b}\right\rangle+\langle a\rangle \\
\beta m\left\langle\frac{b}{a+b}\right\rangle+b      &  -\langle a\rangle-\mu
\end{array}\right).
\end{eqnarray}
When the largest eigenvalue of the Jacobian matrix is null, we can obtain the epidemic threshold 
\begin{subequations}\label{eq10}
\begin{align}
\left(\frac{\beta}{\mu}\right)_c = & G\left(\sqrt{1+H}-1\right),\\
G = & \frac{2\mu+\langle a\rangle+b+\frac{\langle a\rangle}{\left\langle\frac{a}{a+b}\right\rangle}}{2m\mu\left\langle\frac{b}{a+b}\right\rangle},\\
H = & \frac{4\mu(\mu+b+\langle a\rangle)\left\langle\frac{b}{a+b}\right\rangle}{\left\langle\frac{a}{a+b}\right\rangle\left(2\mu+\langle a\rangle+b+\frac{\langle a\rangle}{\left\langle\frac{a}{a+b}\right\rangle}\right)^2}.
\end{align}
\end{subequations}
When the probability distribution $F(x)$ satisfies the Delta distribution, i.e., $F(x)=\delta(x-x_0)$, $G$ and $H$ in Eq.~\eqref{eq10} can be simplified to $G=\frac{(\eta x_0+b)(\mu+\eta x_0+b)}{bm\mu}$ and $H=\frac{b\mu}{\eta x_0(\mu+\eta x_0+b)}$.
It is clear that the epidemic threshold of Eq.~\eqref{eq10} is dependent on $\mu$ on the activity-driven network, which is an important difference from annealed and static limits.
Specifically, as $\mu\gg a$ and $\mu\gg b$, the activity-driven network degenerates into the static network, and the threshold of Eq.~\eqref{eq10} can be rewritten as
\begin{equation}
\left(\frac{\beta}{\mu}\right)_c=\frac{1}{m\left\langle\frac{b}{a+b}\right\rangle}\left(\sqrt{\frac{1}{\left\langle\frac{a}{a+b}\right\rangle}}-1\right).
\end{equation}
Besides, as $\mu\ll a$ and $\mu\ll b$, the activity-driven network will degenerate into the annealed network, for which the threshold of Eq.~\eqref{eq10} approximate 
\begin{equation}
\left(\frac{\beta}{\mu}\right)_c=\frac{1}{m\left(\left\langle\frac{a}{a+b}\right\rangle+\frac{\langle a\rangle}{\langle a\rangle+b}\right)}
\end{equation}
by using Taylor expansion.

\begin{figure}
\includegraphics[width=\linewidth]{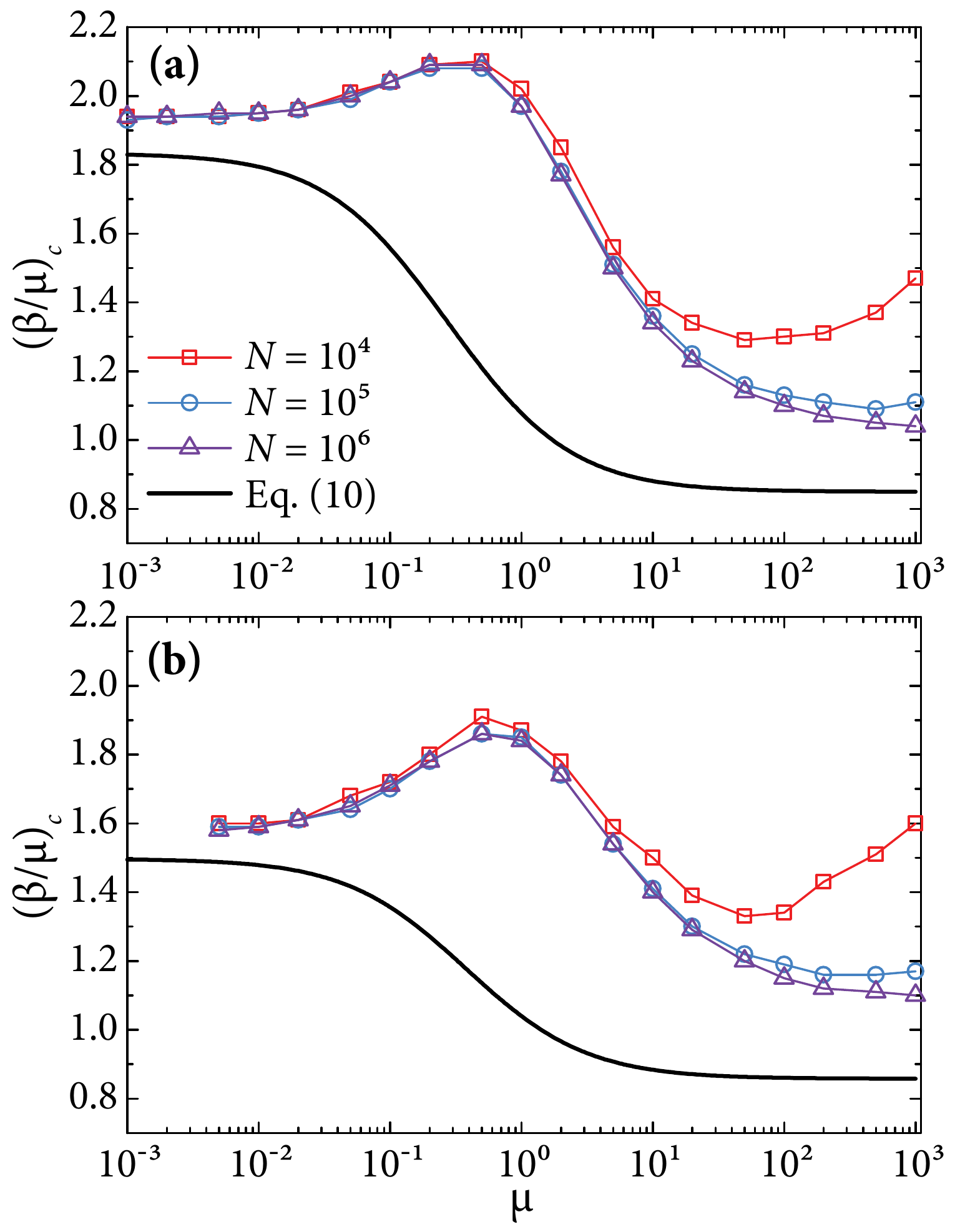}
\caption{The epidemic threshold as a function of recovery rate $\mu$. Parameters: $m=3$, $b=1$; (a) $\eta=1$, $F(x)=\delta(x-0.1)$; (b) $\eta=28$, $F(x)\propto x^{-2.1}$ with $x\in[10^{-3},1]$. The average degrees in (a) or (b) is approximately equal to $0.5$ obtained by Eq.~\eqref{eq5}. Scatters are simulation results with an accuracy of $0.01$. Lines are the theoretical estimations obtained from Eq.~\eqref{eq10} that consider only the network correlations.
\label{fig1}}
\end{figure}

\begin{figure}
\includegraphics[width=\linewidth]{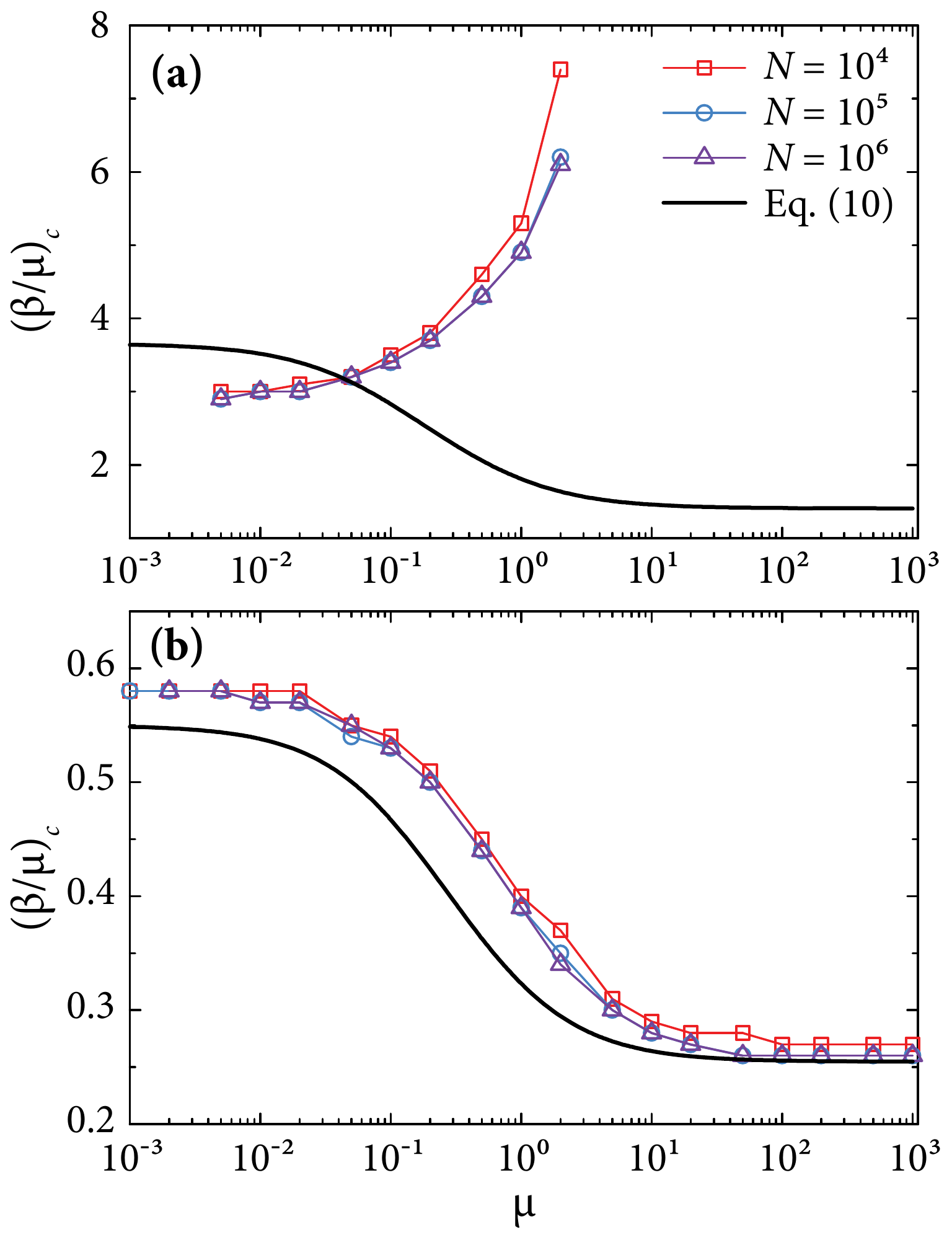}
\caption{The epidemic threshold as a function of recovery rate $\mu$. Parameters:  $b=1$; (a) $m=3$, $\eta=10$, $F(x)\propto x^{-2.1}$ with $x\in[10^{-3},1]$; (b) $m=30$, $\eta=1$, $F(x)=\delta(x-0.1)$. The average degrees in (a) and (b) are approximately equal to $0.23$ and $1.66$, respectively. The accuracy of simulation results of (a) and (b) is $0.1$ and $0.01$, respectively.
\label{fig2}}
\end{figure}

Near the epidemic threshold, the order parameter fluctuates greatly, leading to a critical slowing. The latter means that the relaxation time of the order parameter tends to infinity in the thermodynamic limit.
In general, the internal causes of divergence in relaxation time and divergence in thermodynamic susceptibility are the same.
To estimate the threshold of the SIS model on simulation, a lifetime-based method is proposed in Ref.~\cite{PhysRevLett.111.068701} on static networks, and Ref.~\cite{PhysRevE.96.042310} use this method on activity driven networks.
The lifetime $L$ is defined as the time that passes before the disease either goes extinct or spreads to a finite fraction $C$ of the network, where $C$ is the fraction of distinct nodes ever infected during the simulation.
When $\beta/\mu$ is much less than its critical value $(\beta/\mu)_c$, the disease will quickly die out. If $\beta/\mu \gg (\beta/\mu)_c$, the disease rapidly reaches a steady state.
For $\beta/\mu \sim (\beta/\mu)_c$, we can observe a longer lifetime due to critical slowing down. 
In this case, the opposing mechanisms of infection and spontaneous recovery almost balance out, making spreading dynamics slow.
In the thermodynamic limit, the average lifetime diverges at $(\beta/\mu)_c$ both from below and from above.
For finite systems, we use the maximum value of $L$ to estimate the epidemic threshold.

The simulation results of Fig.~\ref{fig1}, show an interesting nonmonotonic increase of the threshold with $\mu$.
Specifically, the curves show both local maxima and minima, whereas the analytical results are strictly decreasing.
Meanwhile, the threshold tends to have different saturation values in the static or the annealed limits.
The simulation results will increase with the increase of $\mu$ when the analytical results tend to be saturated.
Based on the above analysis, one can learn that network correlation causes the threshold to decrease with the increase of $\mu$, while dynamic correlation has the opposite effect.

We know that the average degree can regulate the strength of the dynamic correlation.
In Fig.~\ref{fig2}(a), we significantly reduce the average degree, at which point the correlation of the dynamic is greatly enhanced and dominant, so we see the threshold monotonically increases as $\mu$ increases.
In contrast, Figure~\ref{fig2}(b) shows that the threshold decreases monotonically as $\mu$ increases, as the average degree increases substantially.
In addition, we can see in Fig.~\ref{fig2}(b) that the results from Eq.~\eqref{eq10} match well with those obtained from Monte Carlo simulations.

\begin{figure}
\includegraphics[width=\linewidth]{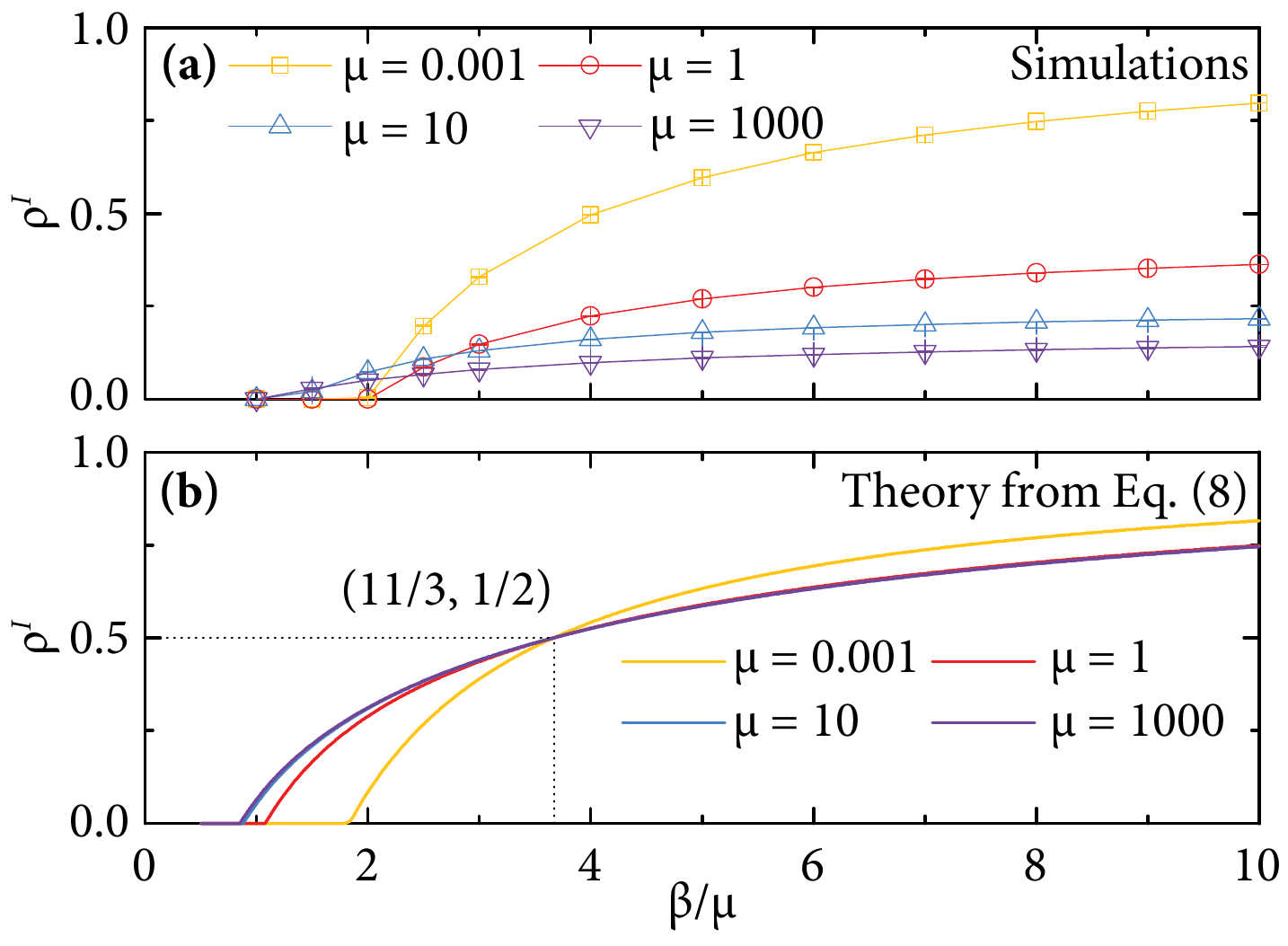}
\caption{The epidemic prevalence $\rho^I$ as a function of $\beta/\mu$ for different recovery rate $\mu$. Parameters: $N=10^5$, $m=3$, $\eta=1$, $F(x)=\delta(x-0.1)$, $b=1$. The scatter plot of panel (a) shows simulation results from averages over 100 independent runs. Lines in panel (b) are the theoretical estimations obtained from Eq.~\eqref{eq8}.
\label{fig3}}
\end{figure}

As a final analysis, we plot the epidemic prevalence $\rho^I$ as a function of $\beta/\mu$ for different $\mu$ in Fig.~\ref{fig3}.
For the analytic results in Fig.~\ref{fig3}(b), we can see that $\rho^I$ appears as a crossing behavior as a function of $\beta/\mu$, which are the results of considering only the network correlation and ignoring the dynamic correlation.
The crossing point occurs precisely at $\rho^I=0.5$ and the corresponding $\beta/\mu=[m\langle a/(a+b)\rangle]^{-1}$.
For the simulation results of Fig.~\ref{fig3}(a), we see that different curves no longer intersect precisely at one point, because the effects of dynamic correlation are not the same for different $\mu$.
Comparing the curves with high $\mu$ in Fig.~\ref{fig3}(a) and Fig.~\ref{fig3}(b), such as $\mu=10$, we can see that $\rho^I$ has a large drop for any $\beta/\mu$, which means that the correlation of dynamic play a key role for high $\mu$.
In addition, we can see in Fig.~\ref{fig3} that our Eq.~\eqref{eq8} can be used to predict the prevalence for low $\mu$.

In summary, we investigated the effects of network correlation and dynamic correlation on SIS epidemics of activity-driven networks.
In particular, we analyze the effects of the two types of correlations in isolation by comparing theoretical and simulation results.
As recovery rate $\mu$ increases, the threshold $(\beta/\mu)_c$ decreases for the effect of network correlation but increases for the effect of dynamic correlation.
Because of this competitive relationship, the curves of the threshold show three types of behavior,
which are monotonically increasing, monotonically decreasing, and first increasing and then decreasing but increasing again.
Meanwhile, the effect of network correlation means that the prevalence produces a crossing behavior, and the effect of dynamic correlation substantially reduces the prevalence.
Finally, our theory can predict the threshold when the average degree $\langle k\rangle$ is high and predict the prevalence for low recovery rate $\mu$. As a final remark, we hope our study will open the door for other analytical studies of the full, two-dimensional parameter space of the SIS model on temporal networks.

\begin{acknowledgments}
    This work was supported by the Shaanxi Fundamental Science Research Project for Mathematics and Physics (Grant No.\ 22JSQ003). PH was supported by JSPS KAKENHI Grant Number JP 21H04595.
\end{acknowledgments}

\appendix

\renewcommand{\theequation}{S\arabic{equation}}
\renewcommand{\thefigure}{S\arabic{figure}}

\section{Simulation procedure}
In this paper, the simulation procedure is roughly divided into two steps. 
The activity-driven network is first evolved to equilibrium. That means that the number of active individuals and the average degree of the network are stable.
Then, the initial infected individuals are randomly selected. The network and dynamics coevolve to dynamic equilibrium. That is, the number of infected individuals reaches a stable level. 
The detailed simulation procedure is as follows:
\begin{enumerate}[label=(\roman*)]
\item Set $\tau=0$, the network starts with $N$ disconnected and inactive nodes.
\item At any time $\tau$, we calculate each individual’s transition rates $\lambda_i(\tau)$. The rate for any inactive individual becoming active is $\lambda_i(\tau)=a_i$. The rate for any active individual becoming inactive is $\lambda_i(\tau)=b$. Summing up all of them, we yield the total transition rate $\omega(\tau)=\sum_i\lambda_i(\tau)$.
\item Time is incremented by $d\tau=1/\omega(\tau)$. The individual whose state is chosen to change at time $\tau+d\tau$ is sampled with a probability proportional to $\lambda_i(\tau)$. The selected individual changes its state.
When the selected individual is activated, it randomly selects $m$ individuals to generate links.
When the selected individual becomes inactive, it deletes all links to itself.
\item Repeat steps (ii)--(iii) until the number of active individuals and the average degree of the network are stable.
\item Set $t=0$, $I_0$ initial infected individuals are randomly selected.
\item At any time $t$, we calculate each individual’s transition rates of the network state $\lambda_1^i(t)$ and those of the dynamic state $\lambda_2^i(t)$. The rate for any inactive individual becoming active is $\lambda_1^i(t)=a_i$. The rate for any active individual becoming inactive is $\lambda_1^i(t)=b$. The rate for any susceptible individual becoming infected is $\lambda_2^i(t)=\beta k_{\text{inf}}$ and $k_{\text{inf}}$ is the number of infected neighbors of the focal individual. The rate for any infected individual recovering is $\lambda_2^i(t)=\mu$. Summing up all of them, we yield the total transition rate $\omega(t)=\sum_i[\lambda_1^i(t)+\lambda_2^i(t)]$.
\item Time is incremented by $dt=1/\omega(t)$. The individual whose state is chosen to change at time $t+dt$ is sampled with a probability proportional to $\lambda_1^i(t)+\lambda_2^i(t)$. 
For the selected individual, one of its network and dynamic states is changed, and the probability sampling is proportional to $\lambda_1^i(t)/[\lambda_1^i(t)+\lambda_2^i(t)]$ and $\lambda_2^i(t)/[\lambda_1^i(t)+\lambda_2^i(t)]$.
When the network state of the selected individual changes, the rules for generating or disconnecting links are the same as those in step (iii).
\item Repeat steps (vi)--(vii) until the number of infected individuals is stable.
\end{enumerate}

\section{Comparison of simulation and theory for average degree}
\begin{SCfigure*}
\includegraphics[width=14cm]{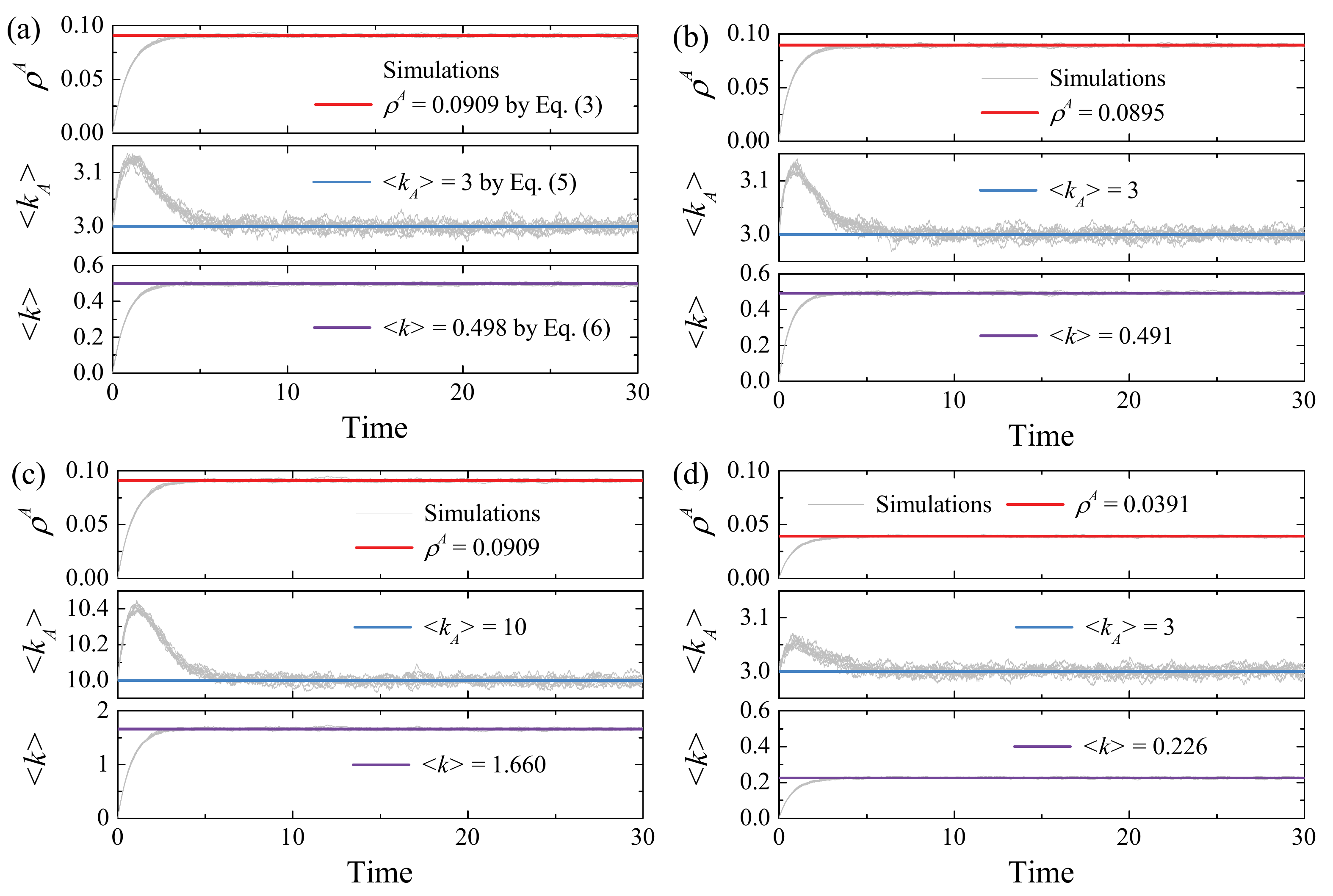}
\caption{
The time series for the fraction of active individuals, the average degree of active individuals, and the average degree of the whole network.
The light gray lines are simulation results of $10$ independent runs.
The thick lines in red, blue, and purple are obtained from Eq.~\eqref{eq2}, Eq.~\eqref{eq4}, and Eq.~\eqref{eq5}, respectively.
Parameters: $N=10^5$, $b=1$; (a) $m=3$, $\eta=1$, $F(x)=\delta(x-0.1)$; (b) $m=3$, $\eta=28$, $F(x)\propto x^{-2.1}$ with $x\in[10^{-3},1]$; (c) $m=10$, $\eta=1$, $F(x)=\delta(x-0.1)$; (d) $m=3$, $\eta=10$, $F(x)\propto x^{-2.1}$ with $x\in[10^{-3},1]$.
\label{s1}}
\end{SCfigure*}

In Fig.~\ref{s1}, we plot the time series of the fraction of active individuals, the average degree of active individuals, and the average degree of the network.
The simulation results, which are denoted by light gray lines, are $10$ independent realizations. The analytical solutions of $\rho^A$, $\langle k_A\rangle$, and $\langle k\rangle$ are denoted by the red, blue, and purple lines, respectively.
It is obvious from Fig.~\ref{s1} that there is an excellent agreement between the analytical solutions and the simulation results.
Since we theoretically consider the correlation of network evolution, this allows the fraction of the average degree of active individuals to be accurately predicted, which provides the basic guarantee for the discussion of coevolution between the network and the dynamic above.

\begin{widetext}

\section{Derivation of Eq.~(8)}
By inserting Eq.~\eqref{eq7} into Eq.~\eqref{eq6}, we have
\begin{subequations}
\begin{eqnarray}\label{eqs1}
\frac{d\rho^{UI_a}}{dt}&=&\beta m\int\rho^{US_a}\rho^{AI_{a'}}da'-(a+\mu)\rho^{UI_a}+b\rho^{AI_a},\\
\frac{d\rho^{AI_a}}{dt}&=&\beta m\int\left(2\rho^{AS_a}\rho^{AI_{a'}}+\rho^{AS_a}\rho^{UI_{a'}}\right)da'+a\rho^{UI_a}-(b+\mu)\rho^{AI_a}.
\end{eqnarray}
\end{subequations}
Combining with $\rho^{US_a}=1-\rho^{A_a}-\rho^{UI_a}$, $\rho^{AS_a}=\rho^{A_a}-\rho^{AI_a}$, and Eq.~\eqref{eq2}, Eq.~\eqref{eqs1} is written as
\begin{subequations}
\begin{eqnarray}\label{eqs2}
\frac{d\rho^{UI_a}}{dt}&=&\beta m\left(\frac{b}{a+b}-\rho^{UI_a}\right)\int\rho^{AI_{a'}}da'-(a+\mu)\rho^{UI_a}+b\rho^{AI_a},\\
\frac{d\rho^{AI_a}}{dt}&=&\beta m\left(\frac{a}{a+b}-\rho^{AI_a}\right)\int\left(2\rho^{AI_{a'}}+\rho^{UI_{a'}}\right)da'+a\rho^{UI_a}-(b+\mu)\rho^{AI_a}.
\end{eqnarray}
\end{subequations}
We define $\rho^{UI}$ and $\rho^{AI}$ as the fraction of individuals in UI state and in AI state, i.e., $\rho^{UI}=\int\rho^{UI_a}da$ and $\rho^{AI}=\int\rho^{AI_a}da$.
By integrating both sides of Eq.~\eqref{eqs2} with respect to $a$, we get
\begin{subequations}
\begin{eqnarray}\label{eqs3}
\frac{d\rho^{UI}}{dt}&=&\beta m\left\langle\frac{b}{a+b}\right\rangle\rho^{AI}-\beta m\rho^{UI}\rho^{AI}-\int a\rho^{UI_a}da+\mu\rho^{UI}+b\rho^{AI},\\
\frac{d\rho^{AI_a}}{dt}&=&\beta m\left(\left\langle\frac{a}{a+b}\right\rangle-\rho^{AI}\right)\left(2\rho^{AI}+\rho^{UI}\right)+\int a\rho^{UI_a}da-(b+\mu)\rho^{AI}.
\end{eqnarray}
\end{subequations}

According to Eq.~\eqref{eq4} and Eq.~\eqref{eq5}, we know that the degree distribution of any class $a$ is the same bimodal distribution. Since we have ignored the correlation of dynamic, see Eq.~\eqref{eq7}, $\rho^{UI_a}$ in steady state is independent of $a$, that is,
\begin{equation}\label{eqs4}
\rho^{UI}=\rho^{UI_a}.
\end{equation}
In addition, we present some simulation results in Fig.~\ref{s2}. Note that the simulation results include the correlation of dynamic. 
As can be seen from Fig.~\ref{s2}(a), for most nodes (unshaded area), there is little difference in $\rho^{UI_a}$ near the threshold even if the dynamic correlation is included.

Finally, by inserting Eq.~\eqref{eqs4} into Eq.~\eqref{eqs3}, we can obtain Eq.~\eqref{eq8}:
\begin{subequations}
\begin{align}
\frac{d\rho^{AI}}{dt} = & \left(2\beta m\left\langle\frac{a}{a+b}\right\rangle-b-\mu\right)\rho^{AI}-2\beta m\left(\rho^{AI}\right)^2+\left(\beta m\left\langle\frac{a}{a+b}\right\rangle+\langle a\rangle\right)\rho^{UI}-\beta m\rho^{UI}\rho^{AI},\\
\frac{d\rho^{UI}}{dt} = & \left(\beta m\left\langle\frac{b}{a+b}\right\rangle+b\right)\rho^{AI}-\left(\langle a\rangle+\mu\right)\rho^{UI}-\beta m\rho^{UI}\rho^{AI}.
\end{align}
\end{subequations}

\begin{SCfigure*}
\includegraphics[width=11cm]{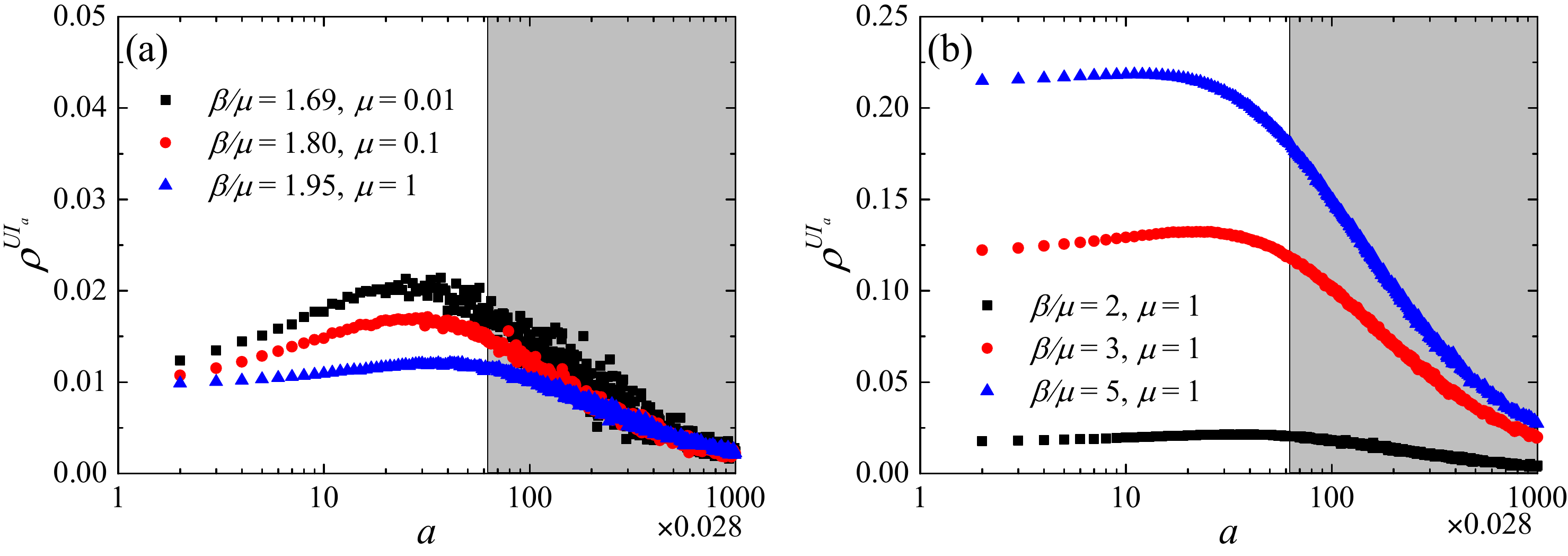}
\caption{
The $\rho^{UI_a}$ as a function of $a$ in the steady state near and far from the epidemic threshold in (a) and (b), respectively.
Parameters: $N=10^5$, $b=1$, $m=3$, $\eta=28$, $F(x)\propto x^{-2.1}$ with $x\in[10^{-3},1]$.
Each simulation runs to $t=5000$. The results represent the average of the last $1000$ time intervals over $50$ independent simulations.
The total fraction of nodes in the shaded area is less than one percent.
\label{s2}}
\end{SCfigure*}

\section{A theoretical solution for the case of 50\% prevalence}
We define $\rho^{I_a}$ ($\rho^{S_a}$) as the fraction of individuals in infected (susceptible) state and class $a$. According to Eq.~\eqref{eqs1}, we have
\begin{subequations}
\begin{align}\label{eqs5}
\frac{d\rho^{I_a}}{dt}= & -\mu\rho^{I_a}+\beta m\int\left(\rho^{US_a}\rho^{AI_{a'}}+2\rho^{AS_a}\rho^{AI_{a'}}+\rho^{AS_a}\rho^{UI_{a'}}\right)da',\\
= & -\mu\rho^{I_a}+\beta m\int\left(\rho^{S_a}\rho^{AI_{a'}}+\rho^{AS_a}\rho^{I_{a'}}\right)da'.
\end{align}
\end{subequations}

We define $\rho^{I}$ and $\rho^{S}$ as the number of individuals in the infected state and in the susceptible state, i.e., $\rho^{I}=\int\rho^{I_a}da$ and $\rho^{S}=\int\rho^{S_a}da$.
By integrating both sides of Eq.~\eqref{eqs5} with respect to $a$, we get
\begin{eqnarray}\label{eqs6}
\frac{d\rho^{I}}{dt}&=&-\mu\rho^{I}+\beta m\left(\rho^{S}\rho^{AI}+\rho^{AS}\rho^{I}\right).
\end{eqnarray}
Consider the special steady state $\rho^{I}=\rho^{S}=0.5$, we have
\begin{equation}\label{eqs7}
\frac{\beta}{\mu}=\frac{1}{m\rho^{A}}=\frac{1}{m\left\langle\frac{a}{a+b}\right\rangle}.
\end{equation}
\end{widetext}

\bibliographystyle{abbrv}
\bibliography{ref}
\end{document}